**Complexity Phenomena and ROMA of the Earth's Magnetospheric Cusp, Hydrodynamic Turbulence, and the Cosmic Web**

[1]Tom Chang, [2]Cheng-chin Wu, [3,4]Marius Echim, [3]Hervé Lamy, [1]Mark Vogelsberger, [5]Lars Hernquist, and [6]Debora Sijacki

**Abstract**

"Dynamic Complexity" is a phenomenon exhibited by a nonlinearly interacting system within which multitudes of different sizes of large scale coherent structures emerge, resulting in a globally nonlinear stochastic behavior vastly different from that could be surmised from the underlying equations of interaction. The hallmark of such nonlinear, complex phenomena is the appearance of intermittent fluctuating events with the mixing and distributions of correlated structures at all scales. We briefly review here a relatively recent method, ROMA (rank-ordered multifractal analysis), explicitly constructed to analyze the intricate details of the distribution and scaling of such types of intermittent structures.

This method is then applied to the analyses of selected examples related to the dynamical plasmas of the cusp region of the Earth's magnetosphere, velocity fluctuations of classical hydrodynamic turbulence, and the distribution of the structures of the cosmic gas obtained through large scale, moving mesh simulations. Differences and similarities of the analyzed results among these complex systems will be contrasted and highlighted.

The first two examples have direct relevance to the Earth's environment (i.e., geoscience) and are summaries of previously reported findings. The third example, though involving phenomena much larger in spatiotemporal scales, with its highly compressible turbulent behavior and the unique simulation technique employed in generating the data, provides direct motivations of applying such analysis to studies of similar multifractal processes in extreme environments of near Earth surroundings. These new results are both exciting and intriguing.

[1] Kavli Institute for Astrophysics and Space Research, Massachusetts Institute of Technology, Cambridge, MA 02139, USA. Email: tsc@space.mit.edu

[2] Institute of Geophysics and Planetary Physics, University of California at Los Angeles, Los Angeles, CA 90095, USA.

[3] Belgian Institute for Space Aeronomy, 1180 Brussels, Belgium.

[4] Institute for Space Sciences, 077125 Bucharest, Romania.

[5] Harvard-Smithsonian Center for Astrophysics, Cambridge, MA 02138, USA.

[6] Institute of Astronomy, University of Cambridge, Cambridge CB3 OHA, UK.

## I. Brief Description of ROMA

**(i) *Preamble.*** Intermittent fluctuating events are popularly analyzed using the structure function and/or partition function methods. These methods investigate the multifractal characteristics of intermittency based on the statistics of the full set of fluctuations. Since most of the observed or simulated intermittent fluctuations are dominated by fluctuations with small amplitudes, the subdominant fractal characteristics of the minority fluctuations—generally of larger amplitudes—are easily masked by those characterized by the dominant population. A new method of rank-ordered multifractal analysis (ROMA) was introduced to specifically address this concern (Chang and Wu, 2008).

**(ii) *Monofractal Behavior.*** Consider, for example, a generic spatial series of certain physical turbulent measure: $\mu(x)$. To address its fluctuating characteristics, it is common to form the scale-dependent difference series $\delta\mu = \mu(x+\delta) - \mu(x)$ and consider the probability distribution functions (PDFs) $P(\delta\mu, \delta)$ for a range of spatial scales $\delta$. Such PDFs for turbulent fluctuations are generally non-Gaussian with extended tails. (See, e.g., Figs. 2, 4, 9, 12, 17-20.) If the phenomenon represented by the fluctuating measure is monofractal, i.e., self-similar, then the scale-dependent PDFs would map onto one scaling function $P_s$ as follows (Chang et al., 2004):

$$P_s(\delta\mu / \delta^s) = \delta^s P(\delta\mu, \delta) \qquad (1)$$

where $s$ is the scaling exponent.

To demonstrate the above assertion, we note that there generally exists an irreducible basis of two independent power-law scale invariants for the variables $(P, \mu, \delta)$: e.g., $\delta\mu / \delta^a = I$ and $P / \delta^b = J$, where $(a, b)$ are the fractal exponents and $(I, J)$ are constants -- i.e. invariants with respect to the scale $\delta$. If the form of $P(\delta\mu, \delta)$ is also invariant as the scale changes, then it has been shown that a functional relation exists between the two invariants (Chang et al., 1973). Imposing the normalization condition for the PDFs, we obtain the one-parameter scaling form as shown in (1), where $s$ is the lone fractal parameter (commonly known as the Hurst exponent). The PDFs are self-similar and monofractal because they map onto one master scaling function $P_s(Y)$ where $Y = \delta\mu / \delta^s$ is a global invariant and $s$ is the only fractal exponent that enters into the scaling expression. Such monofractal mapping was first applied to the analysis of solar wind turbulence by Hnat et al. (2002).

**(iii)** *Structure Functions.* A popular modus operandi designed to study the phenomenon of intermittency is based on the concept of "structure functions", $S_q$, defined by the moments of the PDFs:

$$S_q(\delta\mu, \delta) = \int (\delta\mu)^q P(\delta\mu, \delta) d(\delta\mu) \qquad (2)$$

The motivation here is that different moments emphasize different peaks in the fluctuating series.

Generally, corresponding to each $S_q$ a fractal (structure function) exponent $\zeta_q$ satisfying the power law relation, $S_q = \delta^{\zeta_q}$, for some limited range of small values of $\delta$ may be defined. If $\zeta_q = \zeta_1 q$, then the fractal property of the fluctuating series in that range is characterized by the value of $\zeta_1$. And it may be easily demonstrated that PDFs satisfying the one-parameter scaling form of (1) obey the monofractal property of $\zeta_q = \zeta_1 q$ with $s = \zeta_1$.

When the above linear relation of $\zeta_q$ is violated, the fluctuating phenomenon is considered to be multifractal. Generally, the nonlinear relation between $\zeta_q$ and $q$ is characterized by a noticeable curvature for lower moment orders $q$ and then becomes asymptotically a straight line for large values of $q$. The reason for such a linear asymptotic behavior is due to the unavoidable limitation of available sampling data.

Because the conventional structure function formalism is based on the moments of the full set of fluctuations (which are dominated by those of the small amplitudes), the physical interpretation of the multifractal nature is not easily deciphered by merely examining the curvatures of the deviations from linearity, especially because of their generic linear asymptotic behavior. Furthermore, the structure function exponents are poorly defined because rarely do the actual data exhibit truly power-law relationships between $S_q$ and $\delta$ over the entire scaling range. In addition, even though structure function calculations may be performed conveniently for a fluctuating series for positive values of $q$, they invariably exhibit divergent characteristics for $q < 0$.

**(iv) *ROMA (rank-ordered multifractal analysis).*** Thus, it appears reasonable to search for a procedure that explores the fractal, i.e., power-law, scaling behavior of the subdominant fluctuations by first appropriately isolating out the minority populations and then perform the statistical investigation for each of the isolated populations. Such grouping of fluctuations must depend somehow on the sizes of the fluctuations. However, the groupings cannot depend merely on the raw values of the sizes of the fluctuations because the ranges will be different for different scales. Therefore, we are led to proceed to rank-order the sizes of the fluctuations based on the local invariant $Y = \delta\mu / \delta^s$ where $s$ is the scaling exponent for each (local) grouping.

Consider a differential range of $dY$ in the vicinity of some scaled size $Y = \delta\mu / \delta^s$. We expect the fluctuations whose sizes fall within this differential range to exhibit monofractal behavior characterized by the local scaling exponent $s$ such that the differential structure function $dS_q$ will vary with the scale as $\delta^{sq}$ according to:

$$dS_q \triangleq (\delta\mu)^q P(\delta\mu, \delta) d\delta\mu = \delta^{sq} Y^q P_s(Y) dY \tag{3}$$

Given an ensemble of PDFs $P(\delta\mu, \delta)$, the corresponding multifractal spectrum $s(Y)$ may be obtained approximately (if the ansatz is valid) by integrating the functional differential expression (3) over small contiguous ranges of $\Delta Y$ with the assumption that within each incremental range the scaling exponent $s$ is essentially a constant. Thus, for a range of $\Delta Y$ within $(Y_1, Y_2)$, we form a range-limited structure function as follows:

$$\Delta S_q(\delta\mu, \delta) = \int_{a_1}^{a_2} (\delta\mu)^q P(\delta\mu, \delta) d\delta\mu \simeq \delta^{sq} \int_{Y_1}^{Y_2} Y^q P_s(Y) dY \tag{4}$$

where $a_1 = Y_1 \delta^s$ and $a_2 = Y_2 \delta^s$. We may then search for the value of $s$ such that the scaling property of the range-limited structure function that varies with $s$ is $\Delta S_q(s) \sim \delta^{sq}$. If such a value of $s$ exists, then we have found one region of the multifractal spectrum of the fluctuations such that the PDFs in the range of $\Delta Y$ collapses onto one scaled PDF. Performing this procedure for all contiguous ranges of $\Delta Y$ will produce the approximate rank-ordered multifractal spectrum $s(Y)$ that we are looking for. The determined value of $s$ for each grouping should be un-affected by the statistics of other subsets of fluctuations that are not within the chosen range $\Delta Y$ and therefore should be quantitatively quite accurate. If this spectrum exists, the PDFs for all time lags collapse onto one master multifractal scaled PDF, $P_s(Y)$. The spectrum will be implicit since $Y$ is defined as a function of $s$ (the local Hurst exponent).

The above procedure, commonly known as ROMA, was first introduced by Chang and Wu in 2008. Since then, a flurry of activities in space plasma turbulent studies has utilized this procedure to analyze the multifractal and intermittent characteristics. See, e.g., reviews by Chang et al. (2011) and Chang (2014), and papers by Consolini and De Michelis (2011), Tam et al. (2010) and others.

## II. Turbulent Fluctuations in the Magnetospheric Cusp.

The terrestrial magnetosphere is a bubble in the solar wind carved out by the magnetic field of the Earth. And the cusp is characterized by the magnetospheric region through which plasma from the solar wind can have direct access to the upper ionized atmosphere of the Earth. The plasma in this region is highly turbulent and the intermittent magnetic field fluctuations are statistically anisotropic.

We review below some of the statistical properties of the magnetic energy fluctuations in the cusp region observed by Cluster (a constellation of four identical spacecraft in tetrahedral formation launched in 2000 by the European Space Agency with NASA participation traversing the vicinity and the interior of the Earth's magnetosphere). In space plasmas, physical variables such as the magnetic field intensity are sampled by satellite measurements along the orbit. And observed time ($t$) variations are assumed to be equivalent to spatial variations when the Taylor hypothesis (Taylor, 1938) is valid, i.e.

when a turbulent structure, such as an eddy, transits the spacecraft within a time period smaller than its own time of evolution. Figure 1 displays the time series of magnetic field measurements during a typical cusp passage.

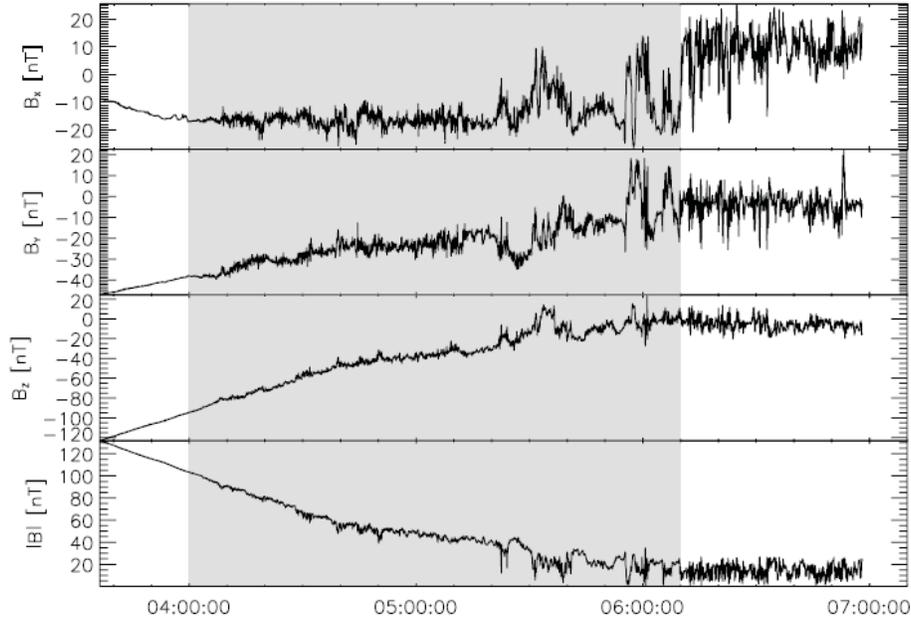

Fig. 1. The components and intensity of the magnetic field measured by Cluster-1. The interval corresponding to the traversal of the cusp region has been shaded in gray. (Echim et al., 2007).

An implicit partial removal of the dipole component of the magnetic field fluctuation data may be achieved by computing first the differences $\delta B^2(t,\tau)$ from the raw data with $\tau$ being the time scale and then the mean value of the fluctuations at each scale is subtracted yielding a new "ensemble" of fluctuations (Echim et al., 2007),

$$\delta b^2 \equiv \frac{\delta B^2(t,\tau) - \left\langle \delta B^2(t,\tau) \right\rangle}{\sigma^2} \tag{5}$$

where the bracket indicates the ensemble average and $\sigma$ is the variance. Typical PDFs were computed for the quantity, $\delta b^2$, where differences $\delta B^2$ have been calculated by moving an overlapping window of width $\tau = 2^j \delta t$ over the entire time interval with $\delta t = 0.0015\,\text{sec}$ being the time resolution of the measurements and $j = 1, 2, ..., 15$, Fig. 2.

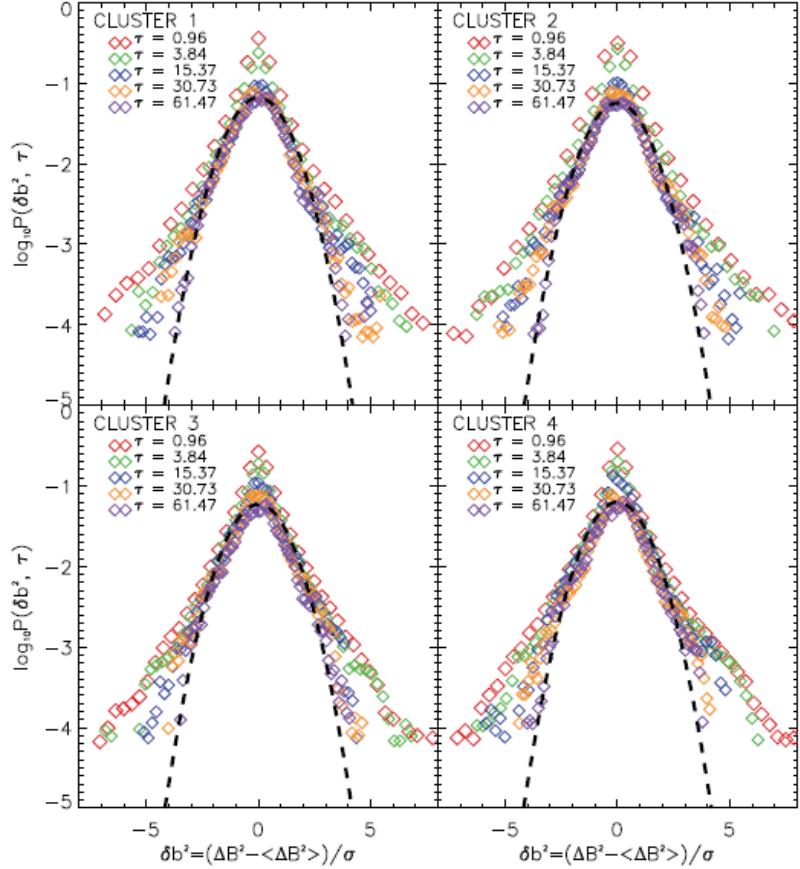

Fig. 2. Typical PDFs of the magnetic energy density fluctuations measured by the four Cluster satellites in the cusp for the time period described in Fig. 1; the PDFs have been scaled with respect to their variance. The scales of $\tau$ are color coded and given in seconds. (Echim et al. 2007).

A ROMA analysis using the afore-mentioned approximate integral technique was performed for the chosen data set in the cusp region, (Lamy et al., 2008, Echim and Lamy, private communication, 2010). We note from Fig. 3 that for small values of $Y$, the fluctuations were persistent *(s > 0.5)*, indicating the turbulence was unstable and probably not yet completely fully developed. For larger values of $Y$, the fluctuations became anti-persistent *(s <0.5)* and the turbulence was probably well-developed and became sparser and sparser as the value of *Y* increased.    (N.B.: The special situation for $s = 0.5$ may be shown to correspond to fluctuations of classical random diffusion.)   The ROMA spectra for all 4 spacecraft were very similar; indicating that despite the magnetic field fluctuations were anisotropic, the magnetic energy density fluctuations in the cusp were essentially statistically isotropic over the distance covered by the cusp passage for spatial scales of the separation distance between the Cluster spacecraft (~1000 km).

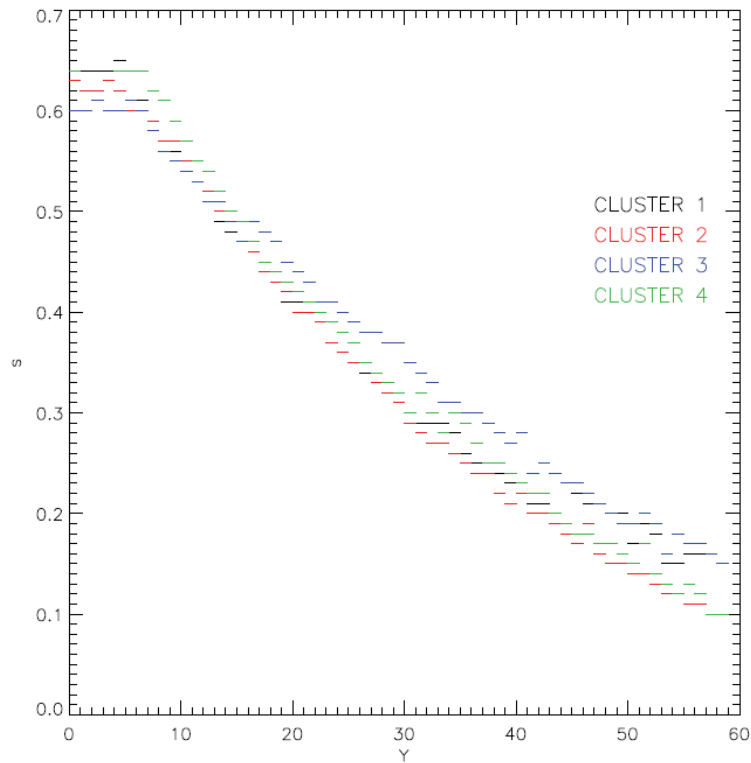

Fig. 3.  Rank-ordered multifractal spectra for the Cluster pass of Fig. 1 in the cusp.  (Echim and Lamy, private communication, 2010; Chang et al., 2010).

### III.  An Interlude – Refined Procedure for ROMA

The above discussion yields ROMA spectra that are step-wise discontinuous. One might wish to improve the calculated result by progressively decreasing the size of $\Delta Y$. However, this procedure may be limited by the statistics within each $\Delta Y$ due to the statistically insufficient amount of available data. Therefore, a refined method of obtaining a continuum of the spectrum $s(Y)$ and the associated scaled PDF $P_s(Y)$ was suggested by Wu and Chang (2011) as described briefly below.

We can write the ROMA scaling relations as:

$$P(X, \delta)\delta^{s(Y)} = P_s(Y) \quad \text{with} \quad Y = X / \delta^{s(Y)} \qquad (6)$$

Thus, for a given value of s, we may plot $P(X, \delta)\delta^{s(Y)}$ against $Y = X / \delta^{s(Y)}$ for the various scales $\delta$ of the PDFs. If the curves intersect at some point $Y_1$ then ROMA is satisfied at $s = s(Y_1)$. On the other hand, if only curves within some range of scales $\delta$ intersect, then ROMA is satisfied at $s = s(Y_1)$ only for that range of scales. (This, in fact, sets the stage for scenarios where there are multiple ROMA scaling ranges as we shall discover below.)

Continue this procedure for the full range of values of $s$ then leads us to a continuum ROMA $s(Y)$ as well as the corresponding scaled PDF $P_s(Y)$. We may then fine tune the result numerically by calculating the PDFs from $s(Y)$ and $P_s(Y)$ and then compare them with the original observed or numerically simulated PDFs. Such a procedure assures that the obtained results are unique. In other words, the continua of $s(Y)$ and $P_s(Y)$ represents the full ensemble of the raw PDFs of the observational or simulated results.

We shall now apply this procedure to the velocity fluctuations of classical driven hydrodynamic turbulence and the complexity phenomenon of the cosmic gas in the following sections.

### IV. Classical Driven Hydrodynamic Turbulence.

It is well-known that fully developed turbulent fluid flows are intermittent and multifractal (Frisch, 1995, and references therein). We have shown that ROMA could be useful in the analysis of fluid turbulence (Chang et al., 2010). We applied the technique to the Johns Hopkins University (JHU) large-scale direct numerical simulation turbulence database based on the Navier-Stokes equations (Perlman et al., 2007; and Li et al., 2008).

Briefly, the data were obtained from a direct numerical simulation of forced isotropic turbulence of a periodic box of $(2\pi)^3$ on a $(1024)^3$ grid using a pseudo-spectral parallel code. Energy was injected by keeping constant the total energy in modes such that their wave-number magnitude is less than or equal to 2. After the simulation reached a statistically stationary state, 1024 frames at every 10 time steps of data, which included the 3 components of the velocity vector and the pressure, were generated and stored into the database. The duration of the stored data was about one large-eddy turnover time of 2.024 . The radial energy spectrum averaged over this duration indicated the existence of an inertial range of wavenumbers approximately between 8 to 60, corresponding to a spatial range from $17\Delta$ to $128\Delta$ where $\Delta = 2\pi / 1024$ is the grid spacing. Instead of the huge $1024^4$ data points, 19 x-planes of data points were used in the analysis. They were arbitrarily selected at various x locations and various times. This set of 19 million data points provided sufficient statistics.

We considered the fluctuations of longitudinal velocity, $\delta v_\parallel$, defined by

$$\delta v_\parallel(\mathbf{r}, \delta) = (\mathbf{v}(\mathbf{r} + \delta\mathbf{i}) - \mathbf{v}(\mathbf{r})) \cdot \mathbf{i} \tag{7}$$

where **i** is the unit vector and $\delta$ is the spatial scale. In the analysis **i** was either in the y- or z-axis and $\delta$ is in the range of $(16\Delta, 160\Delta)$. Figure 4 gives the PDF results for $\delta v_\parallel$ at scale $\delta = 64\Delta$. In computing the PDF, the range of $\delta v_\parallel$ was divided into 1601 bins. For bin number i: $(i - \frac{1}{2})\Delta_\nu < \delta v_\parallel \le (i + \frac{1}{2})\Delta_\nu$. The bin size $\Delta_\nu$ was set as 8/1601 because the maximum value of $|\delta v_\parallel|$ is slightly less than 4. The PDF is asymmetrical in $\delta v_\parallel$ and thus may be decomposed into a symmetrical and an antisymmetric part. The reason for the asymmetry is due to the 3D nature of the fluctuations.

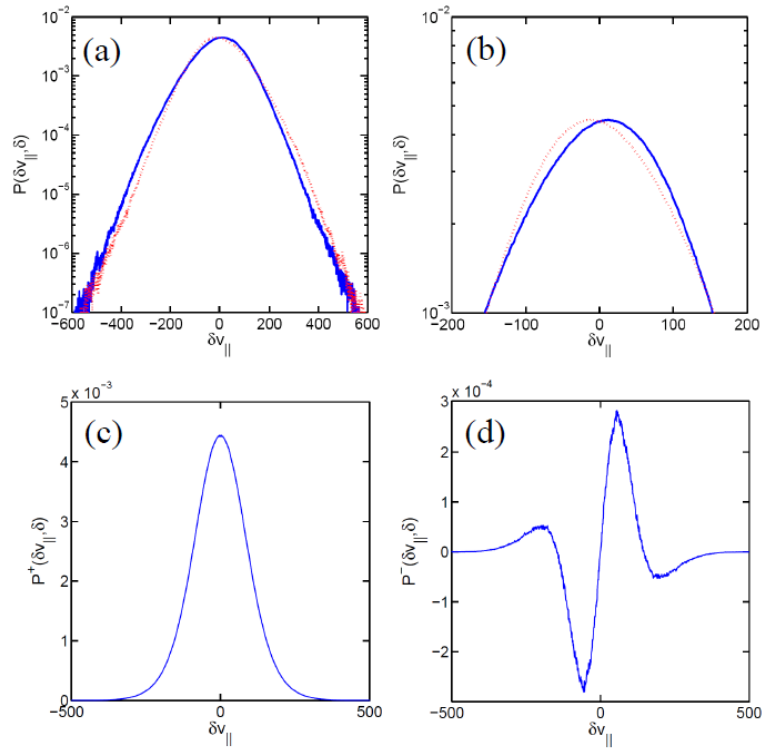

Fig. 4. (a) PDF of $\delta v_\parallel$ for hydrodynamic turbulence at $\delta = 64\Delta$ in units of bin size. (b) is an enlarged version of (a) but with a smaller range of bins. To emphasize the asymmetry, $P(\delta v_\parallel, \delta)$ is shown in solid blue and $P(-\delta v_\parallel, \delta)$ in dashed red. (b) is a zoom of (a). (c) and (d) are the symmetrical and antisymmetric plots $(P^+, P^-)$ of the PDF. (Wu and Chang, 2011).

A ROMA calculation for the PDFs was performed for the simulation results using the refined method as discussed in Sec. 3 (Wu and Chang, 2011). Interestingly, in spite of the asymmetric property of the PDFs, the ROMA spectrum was found to be symmetric as shown in (a) of Fig. 5. The PDFs were mapped onto a scaled master curve $P_s(Y)$ which was asymmetric as shown in (b) and (c) of the figure.

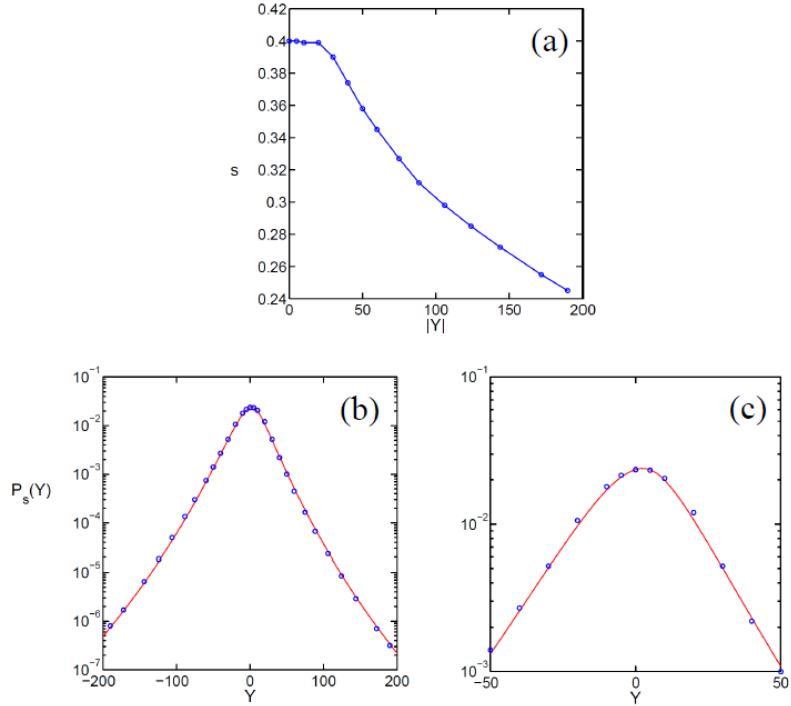

Fig. 5. ROMA spectrum $s(Y)$ and scaled PDF $P_s(Y)$ for the fluctuations of longitudinal velocities for hydrodynamic turbulence. (a) $s(Y)$, which is symmetric with respect to $Y = 0$, shows approximate monofractal behavior at $Y \leq 25$ and decreases monotonically at larger $Y$. (b) $P_s(Y)$ is asymmetric about Y = 0. (c) is a zoom of (b). (Modified from Wu and Chang, 2011).

Figures 6 shows the calculated $P(\delta v_\parallel, \delta)$ based on the ROMA scaling relations of Eq. (6) and the results of $s(Y)$ and $P_s(Y)$ as shown in Fig. 5 for scales from $\delta = 24\Delta$ to $128\Delta$. The comparison with the PDFs from the data is also shown. The results demonstrate that in the inertial range the PDFs of the analyzed fluid turbulence exhibit multifractal scaling that can be described using the ROMA decomposition analysis. Conversely, this also means that the two functions $s(Y)$ and $P_s(Y)$ faithfully reproduce all the PDFs in the inertial range. Thus, the results are unique.

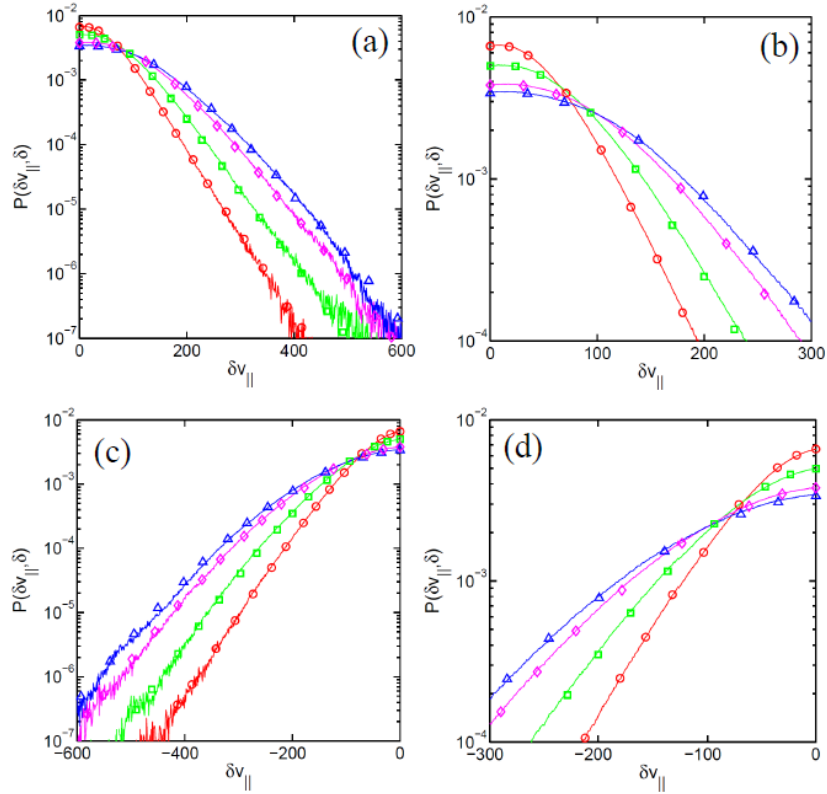

Fig. 6. Plots of $P(\delta v_\parallel, \delta)$ for hydrodynamic turbulence: solid curves are from the simulation data and markers are from ROMA scaling relations: red (circles) for $\delta = 24\Delta$; green (squares) for $\delta = 48\Delta$; magenta (diamonds) for $\delta = 96\Delta$; blue (triangles) for $\delta = 128\Delta$. (a) and (b) are for positive $\delta v_\parallel$; (c) and (d) are for negative $\delta v_\parallel$. (b) and (d) are enlarged plots.

In addition to the fluctuations of the longitudinal velocities, the PDFs of the fluctuations of the square of the velocity, $v^2$, which for incompressible flow are a measure of kinetic energy, also satisfied the ROMA scaling relations. The characteristics of the ROMA spectrum here is very similar to those of the magnetospheric cusp magnetic energy fluctuations, Fig. 7. The difference is that turbulence analyzed here appears to be more well developed and the local Hurst exponent is everywhere antipersistent.

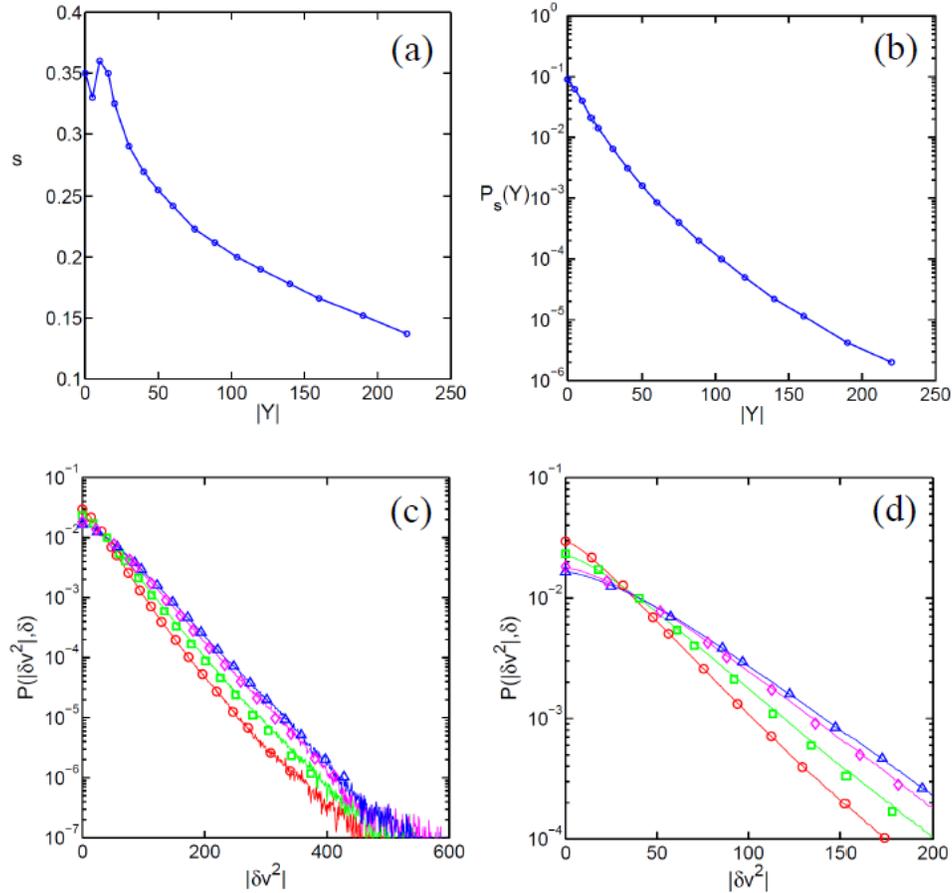

Fig. 7 Fluctuations of $v^2$ for hydrodynamic turbulence. (a) and (b): ROMA spectrum $s(Y)$ and scaled PDF $P_s(Y)$. (c) : plots of $P(\delta v^2, \delta)$ : solid curves are from simulation data and markers are from ROMA scaling relations: red (circles) for $\delta = 24\Delta$; green (squares) for $\delta = 48\Delta$; magenta (diamonds) for $\delta = 96\Delta$; blue (triangles) for $\delta = 128\Delta$. (d) is an enlarged plot for $\delta v^2 \leq 200$.

## V. Unraveling the Complexity of the Cosmic Gas

**(i) *Prologue.*** We are keen to include this third example of ROMA in this treatise both in the spirit of cross-discipline fertilization/exchange of scientific techniques and ideas and in the demonstration of the importance of the role of multiple shock structures in strongly turbulent and compressible gaseous media; situations that may arise in the upper atmosphere and space plasma environments such as the Earth's ionosphere and magnetosphere, the heliosphere and beyond. In addition, the unique moving mesh numerical simulation technique which provided the basic data for the ensuing analysis should prove to be especially useful in the analyses of many realistic domains of the geo/space environments where accurate multi-resolution dynamical studies are required. All the results reported below are new findings.

The commonly accepted theory of cosmic evolution that explains the clumpiness (filaments, pancakes, clusters, voids, etc.) of the baryonic matter content of the Universe is the $\Lambda CDM$ (Lambda cold dark matter) model. It is based on the (FLRW) Friedmann-Lemaître equations and the Robertson-Walker expansion metric of the Einstein's equations of general relativity with the inclusion of the cosmological constant term, $\Lambda$. In addition to the observable baryonic matter, the model includes a cold dark matter component in an effort to explain away the observed anomalous rotational curves of the galaxies and gravitational lensing of light by the clusters of galaxies and to provide the gravitational backbone for the cosmic evolution and structure formation. The cosmological constant (contributing to a constant negative pressure in some form of "dark energy") is included in the model to account for the accelerating expansion of the Universe (Riess et al., 1999; Perlmutter et al., 1999).    For an excellent up-to-date introduction to the subject, see Liddle (2013).

Recent remarkable advances in supercomputing and numerical simulation based on the $\Lambda CDM$ paradigm have provided realistic results that give significant credence to the above theoretical modeling.    The simulations are generally based on the Newtonian approximation of the FLRW equations in terms of the expansion parameter and comoving coordinates (Peebles, 1980).  The physical parameters in the model are guided by the observational inputs and constraints such as those obtained by the WMAP (Wilkinson Microwave Anisotropy Probe) survey and more recently by the Planck survey.

In particular, self-consistent ab initio hydrodynamic simulations of the dark matter and baryonic gas based on the above formalism combined with reasonable feedback mechanisms, radiative cooling, UV ionization and heating, and other relevant physics have yielded reasonable comparisons of the simulated results with observations. See, e.g., Vogelsberger et al. (2013), Torrey et al. (2013) and references contained therein.  For example, there have been studies of the galaxy stellar mass functions, star formation, as well as the comparisons of quasistellar absorption lines associated with the structure of the intergalactic medium such as the Lyman-Alpha forest spectra (Hernquist et al., 1996; Bird et al., 2012 and references contained therein.)

There have been attempts in the statistical studies of the overall density and velocity fields of the cosmic gas, notably those related to the comparisons of observed and simulated power spectra and correlation functions.  There is also, of course, the self-similar spectrum of condensates for the Friedmann cosmology of Press-Schechter (1974) and its comparison with observational and simulation data. In addition, some studies of the intermittent (i.e., non-self-similar) fluctuations of the cosmic structure of the baryonic gas in terms of the traditional methods of structure and partition function analyses based on the simulated results have been reported, (Zhu et al., 2011, and references contained therein).

In the following, we shall address the statistical properties of the density, kinetic energy, velocity and linear momentum of the hierarchical baryonic gas distributions of simulation data in terms of the ROMA technique. As indicated above, such studies will provide the quantitative tool to assess the similarities and differences of intermittent distributions at large, small and intermediate scales. Since ROMA retains all the statistical information of the cosmic distributions, it not only contains the power spectra and correlation information that are usually reported in the observational and simulation literature, but also other quantitative information of the intermittent structures, which are useful for more in-depth comparisons with observations such as those considered by Fang (2006), and Lovejoy et al. ( 2000).

Our ROMA discussed in this paper is based on the recent moving mesh AREPO simulation results of Vogelsberger et al. (2012). AREPO was developed by Springel (2010). It is a second-order accurate finite volume code that employs an unstructured moving mesh based on Voronoi tessellations using a set of moving mesh generating points. The implemented physics for the primordial helium/hydrogen mixture included optically thin radiative cooling, uniform but time dependent ionizing UV background, and simple star formation and supernova feedback (Springel and Hernquist, 2003). The simulation was carried out in a periodic box of 20 $h^{-1}Mpc$ (where $h$ is the Hubble parameter) and the input parameters were chosen to be consistent with the recent WMAP-7 measurements (Komatsu et al., 2011) and other observational constraints. Initial conditions were generated at redshift $z = 99$ based on the spectrum fit of Eisenstein and Hu (1999). The simulation evolved until $z = 0$. Other details may be found from the paper of Vogelsberger et al. (2012).

**(ii)** *Scaling Properties of the Cosmic Density.* The data analyzed in this subsection are taken from the snapshot at $z(redshift) = 0$ of the Vogelsberger et al. simulation (2012) as described in subsection (i). The data given in terms of the moving mesh generating points and Voronoi tessellation were first appropriately redistributed to $512^3$ grid of cubic cells and the fluctuations at different scales were then generated. We note from Fig. 8 that the distribution of the simulated cosmic gas involves densities that span many orders in magnitude. The figure indicates that even for just a two-dimensional slice of the simulation box, the hierarchical structure is already very complicated.

Our analysis, in a sense, is similar to the well-known Press-Schechter (PS) idea (1974) of the mass distribution of the hierarchical condensates for the Friedmann cosmology. Through some heuristic arguments, the PS formalism predicted a power law scaling of the mass condensates and an exponential cutoff with characteristics equivalent to those of the monofractal scaling expression of (1) as discussed in Sec. I. The difference is that our analyses here are focused on the "cosmic gas" and their incremental changes of densities, etc. at different scales. In terms of the cosmic gas, the present results contain more detailed description of its statistical distributions and scaling behavior, particularly those related to their intermittent characteristics.

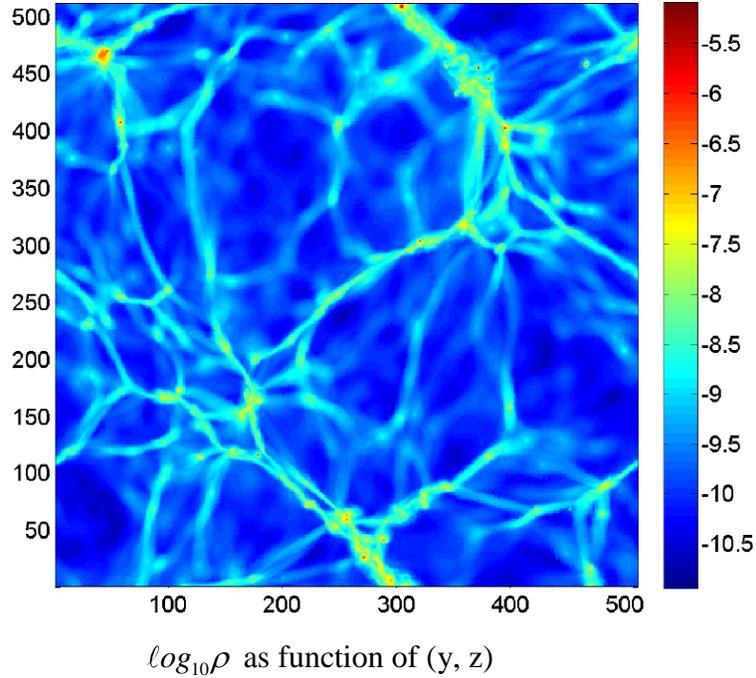

$\ell og_{10}\rho$ as function of (y, z)

Fig. 8. Density distribution $\rho$ of the cosmic gas in units of $10^{10}\odot/(kpc/h)^3$ for $0 < x < \Delta$ with $\Delta = 20/512 h^{-1}Mpc$, at the snapshot of $z(redshift) = 0$.

Figure 9 gives the PDFs for the density fluctuations of the gas at scales $\delta = 32\Delta, 64\Delta, 128\Delta, 256\Delta$ with $\Delta = 20/512 h^{-1}Mpc$. The fluctuation size $|\delta\rho|$ is expressed in units of $(1/800) 10^{10}\odot/\Delta^3$ where $\odot$ is the solar mass. The distributions are distinctly non-Gaussian. We note that at these large scales, the PDFs are scale independent and therefore have asymptotically the same power-law behavior at large values of the fluctuation size $|\delta\rho|$. The scale independent property indicates that the density fluctuations are self-similar and homogeneous for $\delta \geq 32\Delta$.

At smaller scales ($\delta = 4\Delta, 8\Delta, 16\Delta, 32\Delta$), the PDFs become non-self-similar and therefore the fluctuations are intermittent. The calculated ROMA spectrum $s(Y)$ and the scaled PDF $P_s(Y)$ for these scales are given in Fig. 10. The $s(Y)$ is persistent for small local scale invariant $Y$ but soon becomes antipersistent as $Y$ increases; thus, indicating that the fluctuations are predominately fully developed at sufficiently large scaled sizes.

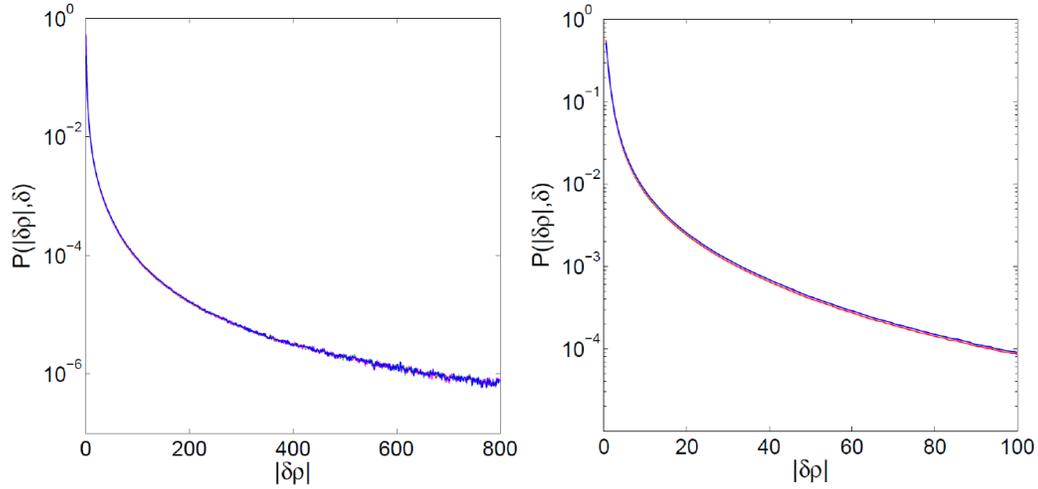

Fig. 9. PDFs of density fluctuations $|\delta\rho|$ of the cosmic gas in units of $(1/800)$ times $10^{10}\odot/\Delta^3$ at $\delta = 32\Delta, 64\Delta, 128\Delta, 256\Delta$ with $\Delta = 20/512 h^{-1} Mpc$. Curves of different colors representing different scales lie on a single curve indicating scale independence. Figure on the right is a zoom of the figure on the left.

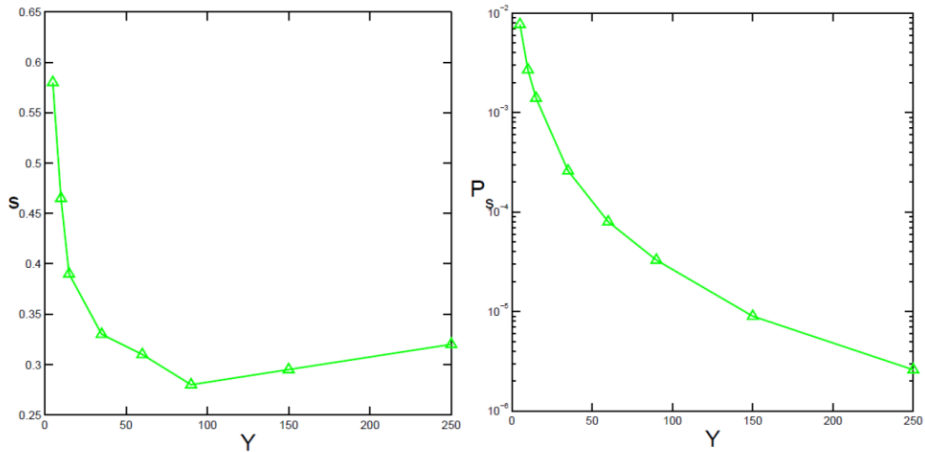

Fig. 10  ROMA scaling for density fluctuations of the cosmic gas at $\delta = 4\Delta, 8\Delta, 16\Delta, 32\Delta$

The comparison of the PDFs generated from ROMA scaling with those of the simulated data is shown in Fig. 11 using the same unit of the bin sizes as Fig. 9. Except for the slight deviations for $\delta = 32\Delta$ the agreements at these small scales are quite striking. We note that the PDFs though non-self-similar nevertheless still exhibit approximately the same power law behavior at large bin sizes.

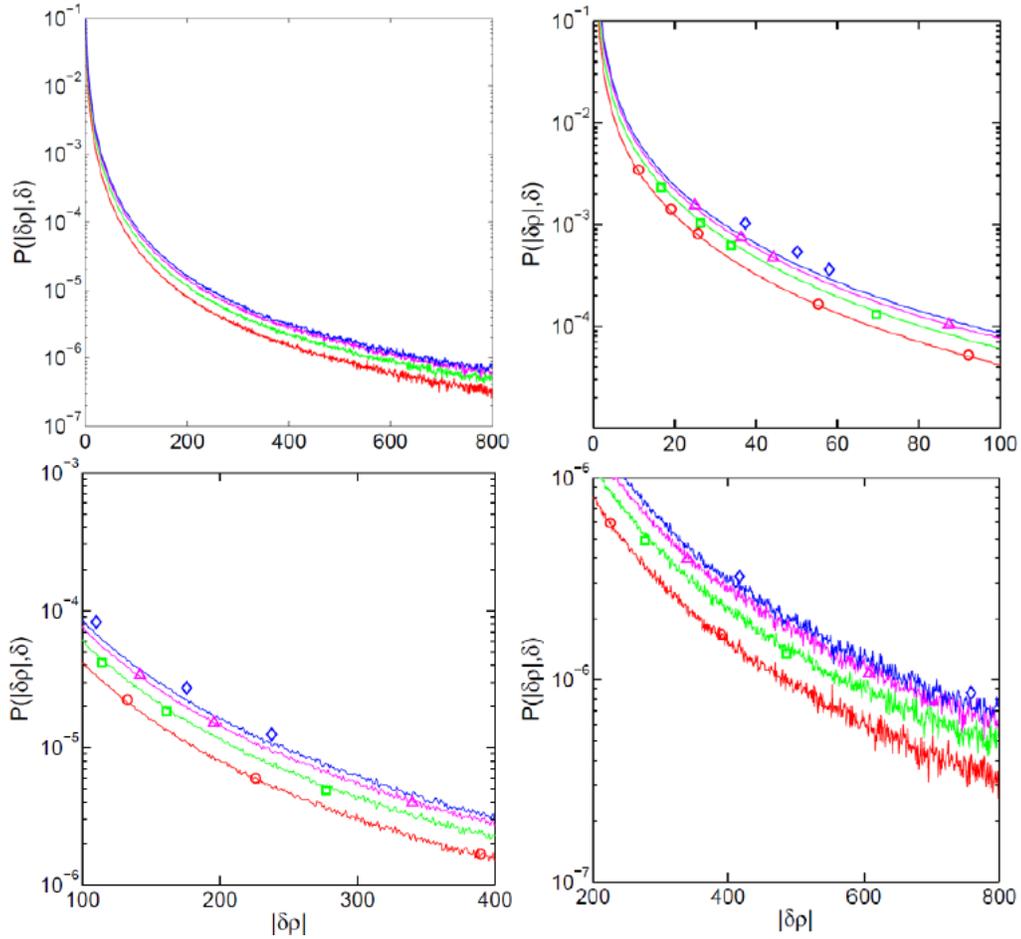

Fig. 11. A sequence of zooms of PDFs for density fluctuations of the cosmic gas generated by ROMA scaling (markers) and simulation data at $\delta = 4\Delta(red), 8\Delta(green), 16\Delta(magenta), 32\Delta(blue)$.

**(iii)** *Asymmetric Intermittency of the Longitudinal Velocity Fluctuations.* Perhaps the most strikingly interesting intermittency scaling behavior of the cosmic gas is related to those characterized by the longitudinal fluctuations $\delta v_{\parallel}$ as defined in Eq. (7). Figure 12 gives PDFs of $\delta v_{\parallel}$ for scales $\delta = 32\Delta, 64\Delta, 96\Delta, 128\Delta$ based on $v_x$, $v_y$, $v_z$, and the average PDF over the three directions for the Vogelsberger (2012) simulation data at $z(redshift) \approx 2.3$. Since the values of $\delta v_{\parallel}$ lie within the range of (944, -1115) km/sec, in our analysis, we set the range to be (-1100, 1100) km/sec, and divide it into 1600 bins, Thus, the units of $\delta v_{\parallel}$ are in 1100/800 km/sec.

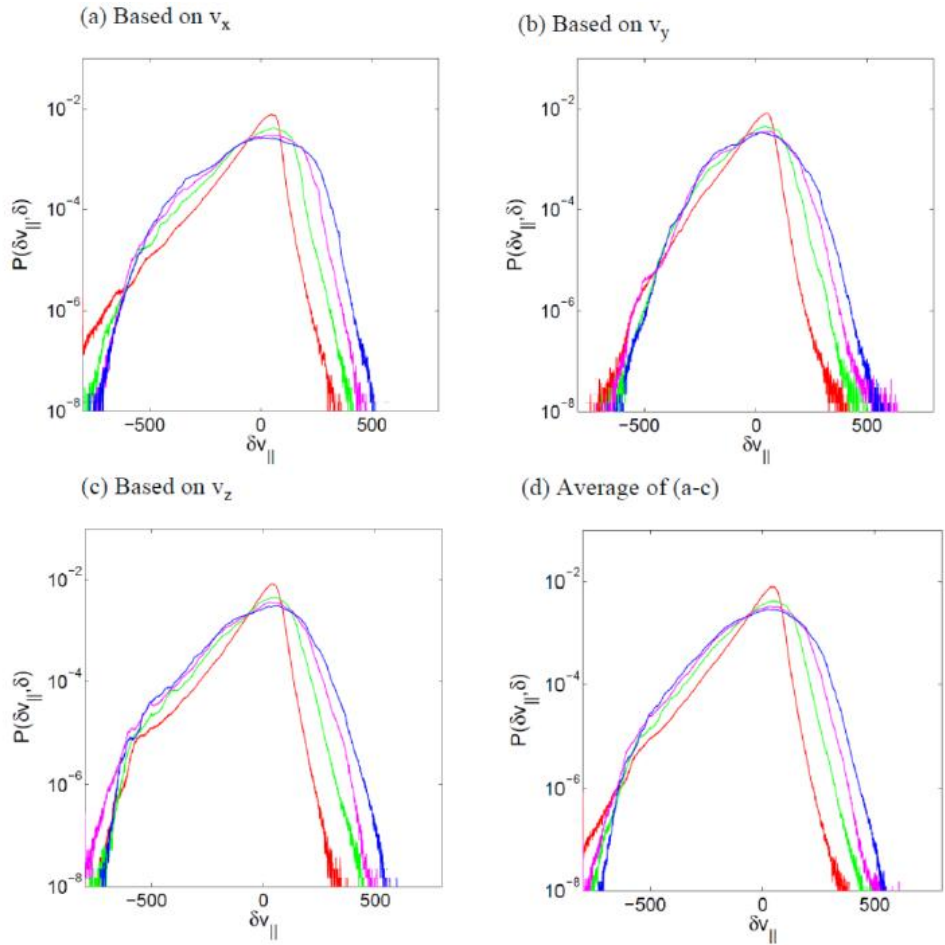

Fig. 12. PDFs of $\delta v_{\parallel}$ of the cosmic gas in units of 1100/800 km/sec.
$\delta = 32\Delta(red), 64\Delta(green), 96\Delta(magenta), 128\Delta(blue)$.

Two properties are apparent from visual inspection of these plots. The first is that the fluctuations are isotropic at these scales. And the second is the asymmetry in $\delta v_\parallel$. The asymmetry is much more pronounced than that of the PDFs for classical driven hydrodynamic turbulence discussed in Sec. IV. The reason may be due to the presence of shock and rarefaction waves in a compressible gas. It is easy to demonstrate that shock waves cause $-\delta v_\parallel$ fluctuations and rarefaction waves create $+\delta v_\parallel$ fluctuations. And these asymmetric contributions to $\delta v_\parallel$ can become quite pronounced for highly compressible media with multiple shock events such as the cosmic gas. Thus, we expect the intermittent fluctuating behavior of $\delta v_\parallel$ of the cosmic gas to be quite different from that for classical driven hydrodynamic turbulence such as that considered in Sec. IV even if both of their traditional structure function spectra exhibit similar nonlinear signatures.

ROMA spectra and scaled PDFs for the PDFs at these large scales are given in Figs. 13 and 14. For positive $\delta v_\parallel$, all PDFs collapse onto one scaled $P_s(Y)$. The corresponding $s(Y)$ is persistent for the small local scaling invariant $Y$ but gradually becomes antipersistent at $Y$ increases, indicating that the fluctuations are somewhat unstable at small sizes but become stable, sparsely distributed, and well developed at large sized fluctuations. For negative $\delta v_\parallel$, the PDFs for $\delta = 32\Delta$ and $64\Delta$ are describable by a scaled PDF with a ROMA spectrum for $-Y$ similar to that for the positive $Y$ values, but at larger scales the PDFs seem to belong to a different scaling category probably due to some unknown physical process(es) and this extraordinary property will require further investigation.

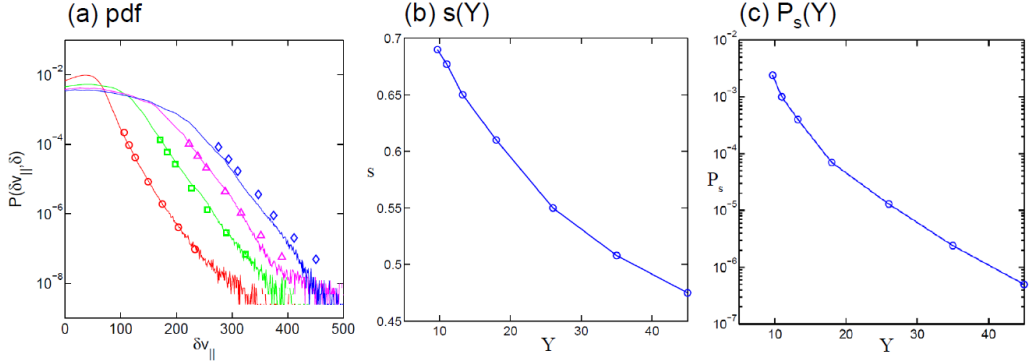

Fig. 13. ROMA scaling and comparison with PDFs of simulation data for $+\delta v_{\parallel}$ at $\delta = 32\Delta(red), 64\Delta(green), 96\Delta(magenta), 128\Delta(blue)$. "$\delta v_{\parallel}$" of the cosmic gas in units of $1100/800$ km/sec. Markers show ROMA scaling using $s(Y)$ and $P_s(Y)$.

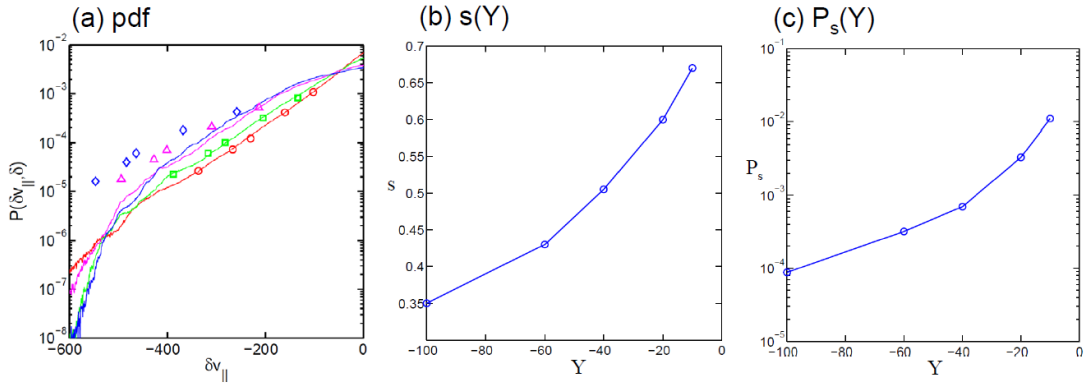

Fig. 14. ROMA scaling and comparison with PDFs of simulation data for $-\delta v_{\parallel}$ at $\delta = 32\Delta(red), 64\Delta(green), 96\Delta(magenta), 128\Delta(blue)$. "$\delta v_{\parallel}$" of the cosmic gas in units of $1100/800$ km/sec. Markers show ROMA scaling using $s(Y)$ and $P_s(Y)$.

For smaller scales, the PDFs map nicely onto separate ROMA scaling curves and the simulated results compare well with the ROMA scaling predictions. Skipping the intermediate details, we summarize the ROMA findings for all scales in Figs. 15 and 16. We note that aside from a restricted large positive $Y$ region, the ROMA results for small scales and large scales fall on separate scaling curves indicating that the physics of the cosmological evolution process are different for scales larger and smaller than a demarcation region in the vicinity of $16\Delta$ to $32\Delta$. AREPO simulation in terms of the mesh generating points and Voronoi cells has the capability of providing multi-resolutions for sparse and dense regions. It will be interesting to utilize this ability to analyze the dense regions at much smaller scales with higher resolution to search for their special scaling properties.

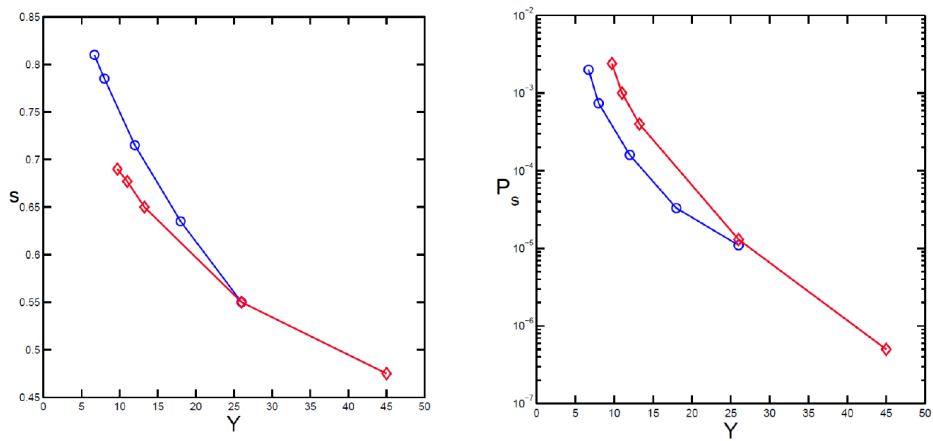

Fig. 15. Combined ROMA $s(+Y)$ and $P_s(+Y)$ of the cosmic gas at small and large scales. Blue circles for $\delta$ between 4 and 32 and Red diamonds for $\delta$ between 32 and 128 in units of $\Delta$.

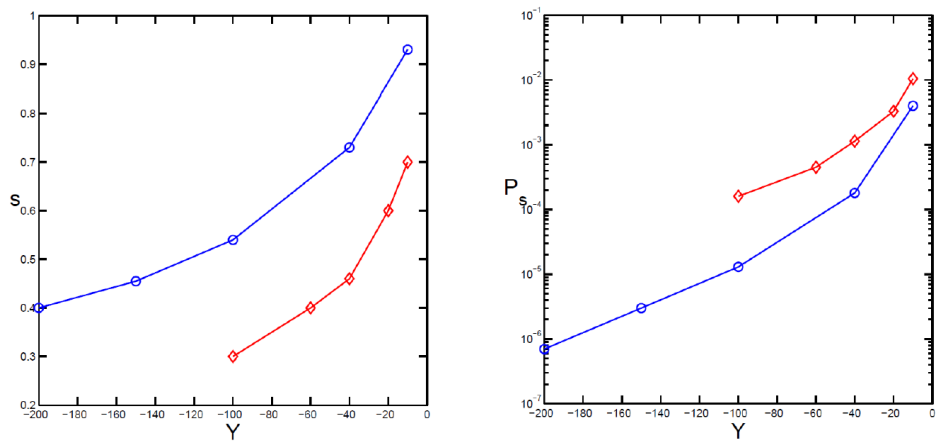

Fig. 16. Combined ROMA $s(-Y)$ and $P_s(-Y)$ of the cosmic gas at small and large scales. Blue circles for $\delta$ between 4 and 16 and Red diamonds for $\delta$ between 32 and 56 in units of $\Delta$.

**(iv).  Fluctuations of *Linear Momentum and Kinetic Energy*.**   In classical driven hydrodynamic turbulence, the longitudinal velocity is a measure for the corresponding linear momentum because the density is a constant.   Since the density also fluctuates intermittently for the compressible cosmic gas, it would be interesting to investigate the behavior of the product of these two entities (i.e., the linear momentum density).   Figure 17 presents the PDFs of the fluctuations of the longitudinal momentum density for $\delta = 32\Delta - 128\Delta$ with $\delta\rho v_\parallel$ in arbitrary units.  The PDFs are strongly non-Gaussian, but essentially scale independent and symmetrical.   There appears only a slight scale dependence at smaller scales of $\delta = 4\Delta - 32\Delta$ (Fig. 18).   And the asymmetry (if any) is noticeable only at small sized fluctuations for all scales.

The reason for this interesting phenomenon may be related to the fact that longitudinal momenta are continuous across both shock and rarefaction waves for compressible flow.   At smaller scales, extra physics related to the real gas effects may influence this property for shock waves and therefore their scaling properties merit further investigation.

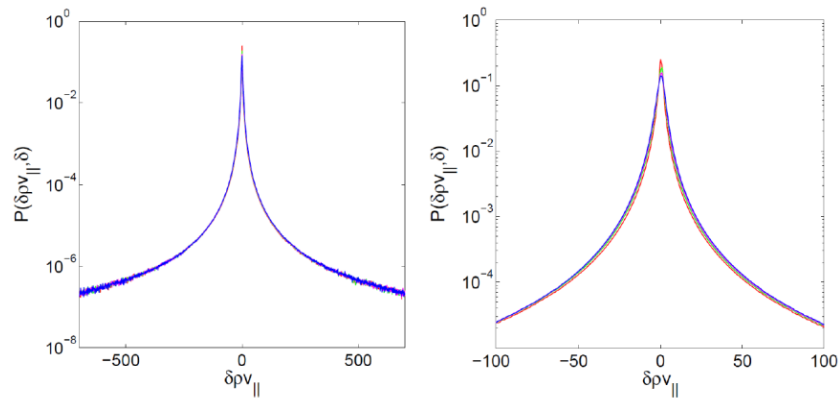

Fig. 17.  PDFs  of  $\delta\rho v_\parallel$ (in arbitrary units) of the cosmic gas  for
$\delta = 32\Delta(red), 64\Delta(green), 96\Delta(magenta), 128\Delta(blue)$ .
Maximum momentum density fluctuations $\sim 320\ 10^{10} \odot (km/s)/\Delta^3$ .

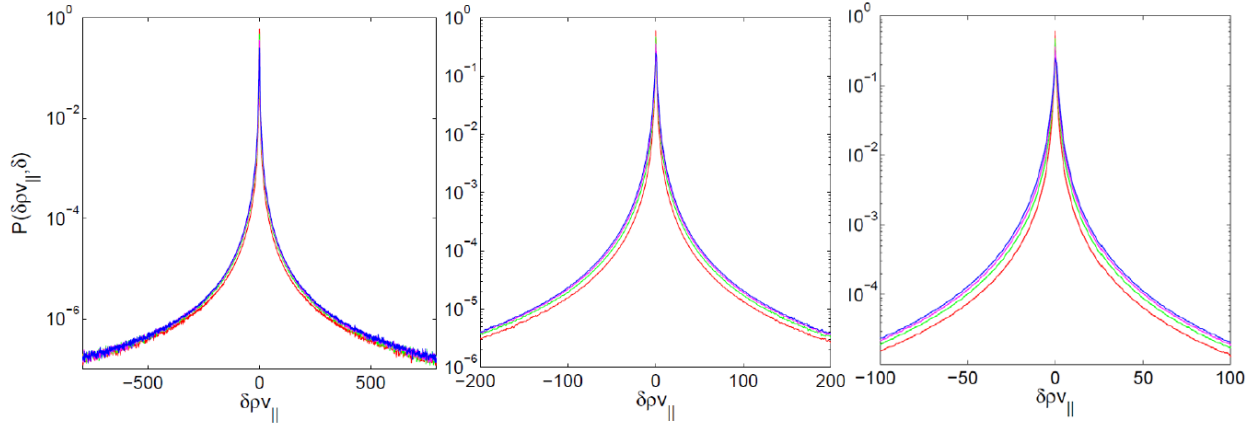

Fig. 18. PDFs of $\delta\rho v_{\parallel}$ (in arbitrary units) of the cosmic gas for $\delta = 4\Delta(red), 8\Delta(green), 16\Delta(magenta), 32\Delta(blue)$.

Also, for classical turbulence, $v^2$ is a measure of kinetic energy. It then provides us a motivation to analyze the scaling behavior of $E = \rho v^2 / 2$, i.e., the kinetic energy density of the cosmic gas. Figure 19 shows that the PDFs of energy density fluctuations for large scales of $\delta = 32\Delta - 128\Delta$ with $\delta E$ in arbitrary units are essentially scale independent and strongly non-Gaussian. In fact, it is only at very small scales, that there is some scale dependence of the PDFs, Fig. 20.

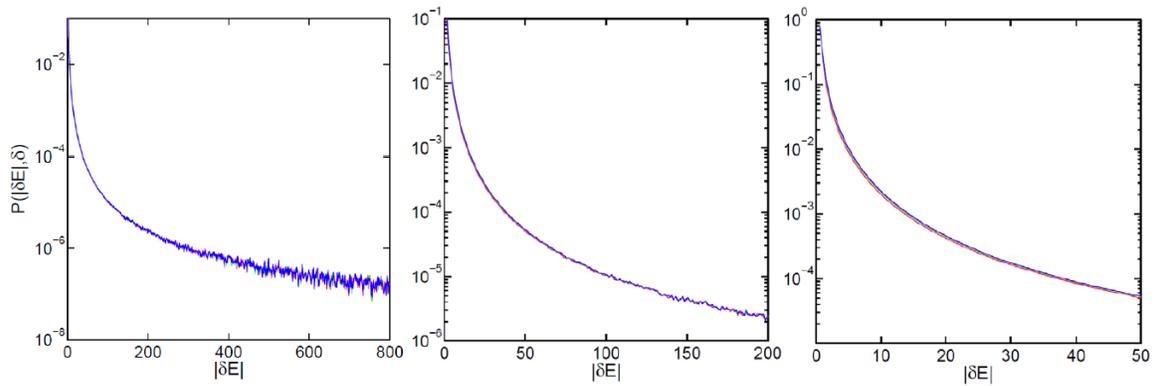

Fig. 19. PDFs of $|\delta E|$ at $\delta = 32\Delta(red), 64\Delta(green), 96\Delta(magenta), 128\Delta(blue)$ for the cosmic gas. $|\delta E|$ in arbitrary units. Maximum energy density fluctuations $\sim 24000$ $10^{10} \odot (km/s)^2 / \Delta^3$.

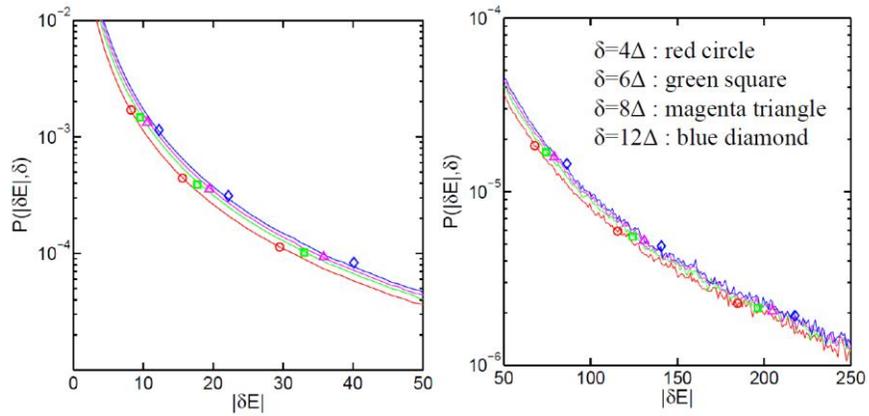

Fig. 20. PDFs of $\left|\delta E\right|$ of the cosmic gas in arbitrary units at small scales. Markers show values obtained by the ROMA scaling relations $s(Y)$, $P_s(Y)$ of Fig. 21.

The ROMA scaling relations, $s(Y)$ and $P_s(Y)$, obtained from the PDFs for small scales (Fig. 21) exhibit nearly the same shapes as those for $\delta v^2$ fluctuations of classical driven hydrodynamic turbulence (Fig. 7). The spectrum $s(Y)$ is everywhere antipersistent and it decreases as $Y$ increases, indicating that the turbulence is fully developed.

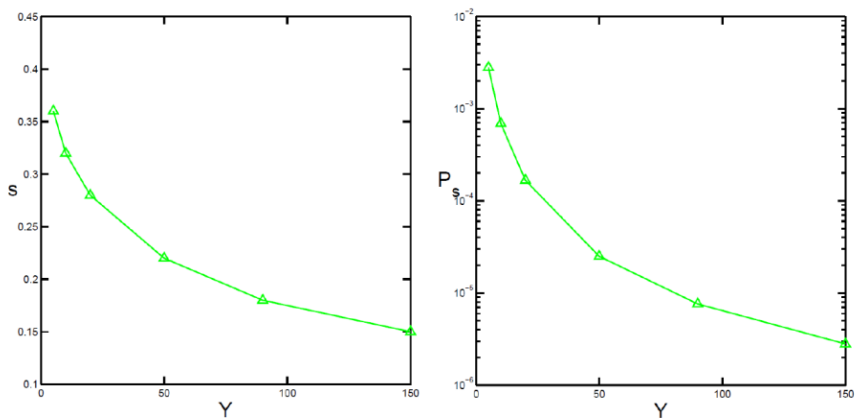

Fig. 21. ROMA scaling relations for the PDFs of $\left|\delta E\right|$ for $\delta = 4\Delta, 6\Delta, 8\Delta, 12\Delta$ for the cosmic gas. Triangles indicate values used for comparison with the simulation data.

## VI. Summary and Conclusions

We provided a brief review of a relatively new method of multifractal analysis (ROMA) designed for in-depth studies of intermittent fluctuations arising from dynamic complexity. Such an implicit multifractal spectrum method has several advantages over the results obtainable using the traditional structure and partition functions. Firstly, the utility of the spectrum is to fully collapse the unscaled PDFs. Secondly, the physical interpretation is clear. It indicates how intermittent the scaled fluctuations are once the spectrum is given. Thirdly, the determination of the values of the fractal nature of the grouped fluctuations is not affected by the statistics of other fluctuations that do not exhibit the same fractal characteristics. Fourthly, the method retains all the statistical information of the analyzed stochastic process and is unique and reversible.

As examples, the method was applied to the analyses of the magnetic energy density fluctuations in the cusp region of the Earth's magnetosphere and classical driven hydrodynamic turbulence.

In addition, for the purpose of relevant cross-discipline fertilization/exchange of scientific techniques and ideas, the paper concludes with a new and exciting analysis of results obtained from an AREPO moving mesh simulation of the cosmic gas. It was discovered that the intermittent characteristics of the longitudinal velocity fluctuations of the gas is vastly different from that generally observed for classical driven hydrodynamic turbulence probably due to the robustly compressible and multiple shock nature of the cosmic baryonic medium. The fluctuations of the longitudinal momentum density of the gas, however, are nearly scale-independent at large scales. In fact, the PDFs for the mass, longitudinal momentum, and kinetic energy densities, though strongly non-Gaussian, are all scale-independent at large scales. At smaller scales, somewhere around and below 1-2 $h^{-1}Mpc$ , the scale-dependence becomes important, indicating special physical effects are beginning to become important in influencing the statistics of the intermittent fluctuations of the cosmic gas.

All the above results indicate that ROMA will be a useful tool in studying and comparing the complexity effects of space plasmas, hydrodynamic turbulence, the climate and geo/space environment, the cosmic web, and other fields of science. A distinct advantage of ROMA is its capability of spotting the similarities and disparities among naturally occurring complexity processes.

**Acknowledgment**

This research is partially supported by the U. S. National Science Foundation and the European Community's Seventh Framework Programme (FP7/ 2007-2013) under Grant agreement no. 313038/STORM. Tom Chang wishes to thank Dr. Diego Perigini for inviting him to present this combined review and report of new findings related to ROMA at the 6[th] International Conference on Fractals and Dynamic Systems in Geoscience in the spirit of providing cross-discipline fertilization/exchange of scientific techniques and ideas in modern fractal analysis.

**References**

Bird, S., Vogelsberger, M., Sijacki, D., Zaldarriaga1, M., Springel, V., and Hernquist, L. (2013), Moving mesh cosmology: properties of neutral hydrogen in absorption, Mon. Not. R. Astron., Soc., 429, 3341.

Chang, T. (2014), *An Introduction to Space Plasma Complexity*, Cambridge University Press, New York, NY.

Chang, T., Wu, C. C., Podesta, Echim, M., Lamy, H., and Tam, S. W. Y. (2010), ROMA (rank-ordered multifractal analyses) of intermittency in space plasmas – a brief tutorial review, Nonlinear Processes in Geophysics, 17, 545.

Chang, T., and Wu, C. C. (2008), Rank-ordered multifractal spectrum for intermittent fluctuations, Phys. Rev. E, 77, 045401(R), doi:10.1103/Phys. Rev. E.77045401.

Chang, T., Tam, S. W. Y., and Wu, C. C. (2004), Complexity induced anisotropic bimodal intermittent turbulence in space plasmas, Phys. Plasmas, 11, 1287.

Chang, T., Hankey, A., and Stanley, H. E. (1973) Generalized scaling hypothesis in multicomponent systems. 1. Classification of critical points by order and scaling at tricriical points, Phys. Rev. B, 8, 346.

Consolini, G. and De Michelis, P. (2011), Rank ordering multifractal analysis of the auroral electrojet index, Nonlinear Processes in Geophysics, 18, 277.

Echim, M. M., Lamy, H., and Chang, T. (2007), Multipoint observations of intermittency in the cusp regions, Nonlinear Processes in Geophysics, 14, 325.

Echim, M, and Lamy, H. (2010), private communication.

Fang, F. (2006), Information of structures in galaxy distribution, Ap. J., 644, 678.

Frisch, U. (1995), *Turbulence*, Cambridge University Press, Cambridge, UK.

Hernquist, L., Katz, N., Weinberg, D. H., and Miralda-Escudé, J. (1996), The Lyman-Alpha forest in the cold dark matter model, Ap. J., 457, L51.

Hnat, B., Chapman, S. C., Rowlands, G., Watkins, N. W., and Farrell, W. M. (2002), Finite size scaling in the solar wind magnetic field energy density as seen by WIND, Geophys. Res. Lett., 29, 1446, 10.1029/2001GL014587.

Komatsu, E., et al. (2011), Seven-year Wilkinson microwave anisotropy probe (WMAP) observations: cosmological interpretation, *Ap. JS,* 192, 18.

Lamy, H., M. Echim, and T. Chang (2008), Rank-ordered multifractal spectrum of intermittent fluctuations in the cusp: a case study with Cluster data, 37th COSPAR Scientific Assembly, Paper D31-0017-08, p. 1686.

Li, Y., Perlman, E., Wan, M., Yang, Y., Burns, R., and Meneveau, C. (2008), A public turbulence database cluster and applications to study Lagrangian evolution of velocity increments in turbulence, J. Turbulence, 9, 1, 2008.

Liddle, A., *An Introduction to Modern Cosmology*, Wiley, Chichester, West Sussex, UK, 2013.

Lovejoy, S., Garrido, P., and Schertzer, D. (2000), Multifractal absolute galactic luminosity distributions and the multifractal Hubble 3/2 law, Physica A, 287, 49.

Peebles, P. (1980), *The large Scale Structure of the Universe*, Princeton University Press, Princeton, NJ.

Perlman, E., Burns, R., Li, Y., and Meneveau, C. (2007), Data exploration of turbulence simulations using a database cluster, Supercomputing SC07, ACM, IEEE, doi:10.1145/1362622, 1362654.

Permutter, S., et al., (1999), Measurements of $\Omega$ and $\Lambda$ from 42 high-redshift supernovae, Ap. J., 517, 565.

Press, W.H., and Schechter, P. (1974), Formation of galaxies and clusters of galaxies by self-similar gravitational condensation, Ap. J., 187, 425.

Riess, a., et al. (1999), The rise time of nearby Type Ia supernovae, Ap. J., 118, 2268.

Springel, V. (2010), E pur si muove: Galiliean-invariant cosmological hydrodynamical simulations on a moving mesh, Mon. Not. R. Astron., Soc., 401, 791.

Springel, V., and Hernquist, L. (2003), The history of star formation in a $\Lambda$ cold dark matter universe, Mon. Not. R. Astron., Soc., 339, 312.

Tam, S. W. Y., et al. (2010), Rank ordered multifractal analysis for intermittent fluctuations with global crossover behavior, Phys. Rev. E, 81, 036404.

Taylor, G. I. (1938), The spectrum of turbulence, Proc. Royal Soc. London A, 164, 476.

Torrey, P., Vogelsberger, M., Genel, S., Sijacki, D., Springel, V., and Hernquist, L. (2013), A physical model for cosmological simulations of galaxy formation: multi-epoch, arXiv:1305.4931v1.

Vogelsberger, M., Sijacki, D., Kereš, D., Springel, V., Hernquist, L. (2012), Moving mesh cosmology: numerical techniques and global statistics, Mon. Not. R. Astron., Soc., 425, 3024.

Vogelsberger, M., Genel, S., Sijacki, D., Torrey, P., Springel, V., and Hernquist, L. (2013), A model for cosmological simulations of galaxy formation physics, arXiv:1305.2913v3.

Wu, C. C., and Chang, T. (2011), Rank-ordered multifractal analysis (ROMA) of probability distributions in fluid turbulence, Nonlinear Processes in Geophysics, 18, 261.

Zhu, W., Feng, L.-L., and Fang, L.Z. (2011), Intermittence of the map of the kinetic Sunyaev-Zel'dovich effect and turbulence of the intergalactic medium, Ap. J. Lett., 734, L14.